\documentclass[12pt,preprint]{aastex}
\newcommand{\gps}{\ensuremath{g_{\rm P1}}}
\newcommand{\rps}{\ensuremath{r_{\rm P1}}}
\newcommand{\ips}{\ensuremath{i_{\rm P1}}}
\newcommand{\zps}{\ensuremath{z_{\rm P1}}}
\newcommand{\yps}{\ensuremath{y_{\rm P1}}}
\newcommand{\wps}{\ensuremath{w_{\rm P1}}}
\newcommand{\grizy}{\gps\rps\ips\zps\yps}
\newcommand{\PS}{\protect \hbox {Pan-STARRS1}}

\shorttitle{The PCS of Pan-STARRS1}
\shortauthors{R.P. Saglia  et al.}
\begin{document}
\title{The Photometric Classification Server for Pan-STARRS1}
%
%
%
\author{
R.P. Saglia\altaffilmark{1,2},
J.L. Tonry\altaffilmark{3},
R. Bender\altaffilmark{1,2},
N. Greisel\altaffilmark{2},
S. Seitz\altaffilmark{1,2},
R. Senger\altaffilmark{1},
J. Snigula\altaffilmark{1},
S. Phleps\altaffilmark{1},
D. Wilman\altaffilmark{1},
C.A.L. Bailer-Jones\altaffilmark{4}, 
R.J. Klement\altaffilmark{4,5},
H.-W. Rix\altaffilmark{4},
K. Smith\altaffilmark{4},
P.J. Green\altaffilmark{6},
W. S. Burgett\altaffilmark{3},
K. C. Chambers\altaffilmark{3}, 
J. N. Heasley\altaffilmark{3},
N. Kaiser\altaffilmark{3},
E. A. Magnier\altaffilmark{3},
J. S. Morgan\altaffilmark{3},
P. A. Price\altaffilmark{7},
C.W. Stubbs\altaffilmark{6},
R. J. Wainscoat\altaffilmark{3}
}

\altaffiltext{1}{Max Planck Institute for Extraterrestrial Physics, Giessenbachstrasse, Postfach 1312, 85741 Garching, Germany}
\altaffiltext{2}{University Observatory Munich, Scheinerstrasse 1, 81679 Munich, Germany}
\altaffiltext{3}{Institute for Astronomy, University of Hawaii, 2680 Woodlawn Drive, Honolulu HI 96822} 
\altaffiltext{4}{Max Planck Institute for Astronomy, K\"onigstuhl 17, D-69117 Heidelberg, Germany} 
\altaffiltext{5}{University of W{\"u}rzburg, Department of Radiation Oncology, D-97080 W{\"u}rzburg, Germany} 
\altaffiltext{6}{Harvard-Smithsonian Center for Astrophysics, 60 Garden Street, Cambridge, MA 02138} 
\altaffiltext{7}{Department of Astrophysical Sciences, Princeton University, Princeton, NJ 08544, USA} 
\begin{abstract}
  The Pan-STARRS1 survey is obtaining multi-epoch imaging in 5 bands
  (\gps \rps \ips \zps \yps) over the entire sky North of declination
  $-30\deg$. We describe here the implementation of the Photometric
  Classification Server (PCS) for \PS. PCS will allow the automatic
  classification of objects into star/galaxy/quasar classes based on
  colors, the measurement of photometric redshifts for extragalactic
  objects, and constrain stellar parameters for stellar objects,
  working at the catalog level.  We present tests of the system based
  on high signal-to-noise photometry derived from the Medium Deep
  Fields of \PS, using available spectroscopic surveys as training
  and/or verification sets. We show that the \PS\ photometry delivers
  classifications and photometric redshifts as good as the Sloan
  Digital Sky Survey (SDSS) photometry to the same magnitude limits.
  In particular, our preliminary results, based on this relatively
  limited dataset down to the SDSS spectroscopic limits and therefore
  potentially improvable, show that stars are correctly classified as
  such in 85\% of cases, galaxies in 97\% and QSOs in 84\%. False
    positives are less than 1\% for galaxies, $\approx 19$\% for stars
    and $\approx 28$\% for QSOs. Moreover, photometric redshifts for
  1000 luminous red galaxies up to redshift 0.5 are determined to
  2.4\% precision (defined as $1.48\times
  Median|z_{phot}-z_{spec}|/(1+z)$) with just 0.4\% catastrophic
  outliers and small (-0.5\%) residual bias.  For bluer galaxies up to
  the same redshift the residual bias (on average -0.5\%) trend,
  percentage of catastrophic failures (1.2\%) and precision (4.2\%)
  are higher, but still interestingly small for many science
  applications. Good photometric redshifts (to 5\%) can be obtained
  for at most 60\% of the QSOs of the sample. PCS will create a value
  added catalog with classifications and photometric redshifts for
  eventually many millions sources.
\end{abstract}


\vfil
\eject
\clearpage

\section{INTRODUCTION}
\label{sec:intro}

Pan-STARRS1, the prototype of the Panoramic Survey Telescope and Rapid
Response System, started scientific survey operations in May 2010 and
is producing a 5 band (\gps \rps \ips \-\zps \yps) imaging for 3/4 of
the sky that will be $\approx 1$ mag deeper than the Sloan Digital Sky
Survey  \citep[SDSS,][]{York00,SDSS} at the end of the foreseen 3 years
of operations. Approximately two hundred million galaxies, a similar
number of stars, about a million quasars and $\approx 7000$ Type Ia
Supernovae will be detected. In addition to the search for so-called
``killer asteroids'', the science cases driving \PS\ are both galactic
and extragalactic.  Extragalactic goals range from Baryonic Acoustic
Oscillations and growth of structure, to weak shear, galaxy-galaxy
lensing and lensing tomography. All rely on the determination of
accurate photometric redshifts for extremely large numbers of
galaxies.  The galactic science cases focus on the search for very
cool stars and the structure of the Milky Way, requiring good
star/galaxy (photometric) classification and constraints on stellar
parameters. Further science goals, such as the detection of high
redshift quasars and galaxies, quasar/quasar and quasar/galaxy
  clustering, or the study of how galaxies evolve with cosmic time
profit from the availability of good photometric redshifts and
star/galaxy photometric classification. Therefore in the last years we
have designed \citep{Saglia08, Snigula09} and implemented the
Photometric Classification Server (PCS) for \PS\ to derive and
administrate photometric redshift estimates and probability
distributions; star/galaxy classification and stellar parameters for
extremely large datasets. The present paper describes the system and
its performances in various tests.

The structure of the paper is as follows. The \PS\ system is
sketched out in Section \ref{sec:system}, the observations we used
are described in Section \ref{sec:observations} and data processing is
outlined in Section \ref{sec:processing}.  Section
\ref{sec:description} presents the PCS system, its algorithms and
components, and the implementation. Section \ref{sec:results}
discusses the tests of the system, which is followed by our conclusions in
Section \ref{sec:conclusions}.


\section{The Pan-STARRS1 Telescope, Camera, and Image Processing}
\label{sec:system}

The \PS\ system is a high-etendue wide-field imaging system, designed
for dedicated survey operations. The system is installed on the peak
of Haleakala on the island of Maui in the Hawaiian island
chain. Routine observations are conducted remotely, from the Advanced
Technology research Center in Pukalani.  We provide below a terse
summary of the Pan-STARRS1 survey instrumentation.  A more complete
description of the Pan-STARRS1 system, both hardware and software, is
provided by \cite{PS1_system}. The survey philosophy and execution
strategy are described by \cite{PS_MDRM}.

The \PS\ optical design \citep{PS1_optics} uses a 1.8~meter diameter
primary mirror, and a 0.9~m secondary.  The resulting converging beam
then passes through two refractive correctors, a
$48$~cm~$\times~48$~cm interference filter, and a final refractive
corrector that is the dewar window. The entire optical system delivers
an $f$/4.4 beam and an image with a diameter of 3.3 degrees, with low
distortion.  The Pan-STARRS1 imager \citep{PS1_GPCA} comprises a total
of 60 detectors, with $4800\times4800$ 10~$\mu$m pixels that each
subtend 0.258~arcsec.  The detectors are back-illuminated CCDs,
manufactured by Lincoln Laboratory, and read out using a StarGrasp CCD
controller, with a readout time of 7 seconds for a full unbinned
image. Initial performance assessments are presented by
\cite{PS1_GPCB}.

The \PS\ observations are obtained through a set of five broadband
filters, which we have designated as \gps, \rps, \ips, \zps, and
\yps. Under certain circumstances \PS\ observations are obtained with
a sixth, ``wide'' filter designated as \wps\ that essentially spans
\gps, \rps, and \ips.  Although the filter system for \PS\ has much in
common with that used in previous surveys, such as the SDSS, there are
important differences. The \gps\ filter extends 20~nm redward of
$g_{SDSS}$, paying the price of 5577\AA\ sky emission for greater
sensitivity and lower systematics for photometric redshifts, and the
\zps\ filter is cut off at 930~nm, giving it a different response from
$z_{SDSS}$.  SDSS has no corresponding \yps\ filter, while \PS\ is
lacking u-band photometry that SDSS provides.  Further information on
the passband shapes is described in \cite{PS_lasercal}.  Provisional
response functions (including 1.3 airmasses of atmosphere) are
available at the project's web site
\footnote[1]{http://svn.pan-starrs.ifa.hawaii.edu/trac/ipp/wiki/PS1\_Photometric\_System}.
Photometry is in the ``natural'' \PS\ system, $m=-2.5log(flux)+m'$,
with a single zeropoint adjustment $m'$ made in each band to conform
to the AB magnitude scale \citep{JTphoto}.  \PS\ magnitudes are
interpreted as being at the top of the atmosphere, with 1.3 airmasses
of atmospheric attenuation being included in the system response
function. No correction for Galactic extinction is applied to the \PS\
magnitudes.  We stress that, like SDSS, \PS\ uses the AB photometric
system and there is no arbitrariness in the definition. Flux
representations are limited only by how accurately we know the system
response function vs. wavelength.

Images obtained by the Pan-STARRS1 system are processed through the
Image Processing Pipeline (IPP) \citep{PS1_IPP}, on a computer cluster
at the Maui High Performance Computer Center. The pipeline runs the
images through a succession of stages, including flat-fielding
(``de-trending''), a flux-conserving warping to a sky-based image
plane, masking and artifact removal, and object detection and
photometry. The IPP also performs image subtraction to allow for the
prompt detection of variables and transient phenomena. Mask and
variance arrays are carried forward at each stage of the IPP
processing. Photometric and astrometric measurements performed by the
IPP system are described in \cite{PS1_photometry} and
\cite{PS1_astrometry} respectively.

The details of the photometric calibration and the \PS\ zeropoint
scale will be presented in a subsequent publication \citep{JTphoto},
and \cite{EMphoto} will provide the application to a consistent
photometric catalog over the 3/4 sky observed by \PS.  


\section{Observations}
\label{sec:observations}

This paper uses images and photometry from the Pan-STARRS1 Medium-Deep
Field survey. In addition to covering the sky at $\delta>-30\deg$ in 5
bands, the Pan-STARRS1 survey has obtained deeper multi-epoch images
in the \PS\ \gps, \rps, \ips, \zps\ and \yps\ bands of the fields
listed in Table \ref{table:fields}. The typical Medium-deep cadence of
observations is $8\times113$ s in the \gps\ and \rps\ bands the first
night, $8\times240$ s in the \ips\ band the second night, $8\times240$
s in the \zps\ band the third night, $8\times113$ s in the \gps\ and
\rps\ bands in the forth night, and on each of the 3 nights on either
side of Full Moon $8\times240$ s in the \yps\ band. The 5 $\sigma$
point source detection limits achieved in the various \grizy\ bands,
as well as other statistics of potential interest, are presented in
Table \ref{table:depths} for the co-added stacks. They represent the
depth of stacks at the time of writing, as observations are still
on-going. In the following we will only consider the fields MDF03 to
10, that overlap with SDSS.


\begin{table}[htdp]
\caption{\PS\ Medium-Deep Field Centers. }
\begin{center}
\begin{tabular}{lrcr}
\hline
\hline
Field & Alternative names & RA (degrees, J2000) & Dec (degrees, J2000) \\
\hline
MDF01  & MD01, PS1-MD01 & ~35.875 & $  4.250$~~~~~~~~~~~~ \\
MDF02  & MD02, PS1-MD02 & ~53.100 & $-27.800$~~~~~~~~~~~~ \\
MDF03  & MD03, PS1-MD03 & 130.592 & $ 44.317$~~~~~~~~~~~~ \\
MDF04  & MD04, PS1-MD04 & 150.000 & $  2.200$~~~~~~~~~~~~ \\
MDF05  & MD05, PS1-MD05 & 161.917 & $ 58.083$~~~~~~~~~~~~ \\
MDF06  & MD06, PS1-MD06 & 185.000 & $ 47.117$~~~~~~~~~~~~ \\
MDF07  & MD07, PS1-MD07 & 213.704 & $ 53.083$~~~~~~~~~~~~ \\
MDF08  & MD08, PS1-MD08 & 242.787 & $ 54.950$~~~~~~~~~~~~ \\
MDF09  & MD09, PS1-MD09 & 334.188 & $  0.283$~~~~~~~~~~~~ \\
MDF10  & MD10, PS1-MD10 & 352.312 & $ -0.433$~~~~~~~~~~~~ \\
\hline
\end{tabular}
\end{center}
\label{table:fields}
\end{table}


\begin{table}[htdp]
\caption{\PS\ MDF Statistics, Apr 2009--Apr 2011.}
\begin{center}
\begin{tabular}{lcrcccc|lcrcccc}
\hline
\hline
Field&Filter& $N$ & $\log t$ & $PSF$ & $\langle w\rangle$ & $m_{lim}$ &
Field&Filter& $N$ & $\log t$ & $PSF$ & $\langle w\rangle$ & $m_{lim}$\\
\hline
MDF01 & \gps & 42 & 4.7 & 1.25 & 1.55 & 24.5 & MDF06 & \gps & 38 & 4.6 & 1.25 & 1.56 & 24.4\\
MDF01 & \rps & 42 & 4.7 & 1.15 & 1.35 & 24.4 & MDF06 & \rps & 39 & 4.6 & 1.18 & 1.45 & 24.2\\
MDF01 & \ips & 41 & 4.9 & 1.05 & 1.27 & 24.4 & MDF06 & \ips & 41 & 4.9 & 1.14 & 1.39 & 24.3\\
MDF01 & \zps & 41 & 4.9 & 1.03 & 1.24 & 23.9 & MDF06 & \zps & 38 & 4.9 & 1.05 & 1.30 & 23.7\\
MDF01 & \yps & 21 & 4.6 & 0.95 & 1.17 & 22.4 & MDF06 & \yps & 24 & 4.7 & 1.00 & 1.25 & 22.4\\
MDF02 & \gps & 30 & 4.5 & 1.31 & 1.79 & 24.2 & MDF07 & \gps & 36 & 4.5 & 1.23 & 1.68 & 24.3\\
MDF02 & \rps & 29 & 4.5 & 1.20 & 1.74 & 24.1 & MDF07 & \rps & 39 & 4.5 & 1.13 & 1.46 & 24.2\\
MDF02 & \ips & 30 & 4.8 & 1.11 & 1.50 & 24.2 & MDF07 & \ips & 39 & 4.9 & 1.14 & 1.44 & 24.2\\
MDF02 & \zps & 33 & 4.8 & 1.06 & 1.30 & 23.6 & MDF07 & \zps & 43 & 4.9 & 1.08 & 1.37 & 23.7\\
MDF02 & \yps & 16 & 4.5 & 1.14 & 1.42 & 22.1 & MDF07 & \yps & 30 & 4.8 & 1.01 & 1.28 & 22.5\\
MDF03 & \gps & 38 & 4.6 & 1.18 & 1.44 & 24.5 & MDF08 & \gps & 38 & 4.5 & 1.27 & 1.68 & 24.3\\
MDF03 & \rps & 37 & 4.6 & 1.09 & 1.28 & 24.4 & MDF08 & \rps & 38 & 4.5 & 1.14 & 1.47 & 24.2\\
MDF03 & \ips & 41 & 4.9 & 1.06 & 1.31 & 24.4 & MDF08 & \ips & 33 & 4.8 & 1.07 & 1.34 & 24.2\\
MDF03 & \zps & 42 & 5.0 & 1.03 & 1.27 & 23.9 & MDF08 & \zps & 40 & 4.9 & 1.09 & 1.39 & 23.7\\
MDF03 & \yps & 20 & 4.6 & 1.00 & 1.36 & 22.4 & MDF08 & \yps & 32 & 4.9 & 0.98 & 1.27 & 22.7\\
MDF04 & \gps & 35 & 4.6 & 1.17 & 1.52 & 24.5 & MDF09 & \gps & 34 & 4.5 & 1.26 & 1.55 & 24.3\\
MDF04 & \rps & 37 & 4.6 & 1.09 & 1.46 & 24.3 & MDF09 & \rps & 33 & 4.5 & 1.15 & 1.42 & 24.1\\
MDF04 & \ips & 35 & 4.9 & 1.07 & 1.35 & 24.3 & MDF09 & \ips & 34 & 4.8 & 1.02 & 1.36 & 24.3\\
MDF04 & \zps & 28 & 4.8 & 1.03 & 1.32 & 23.6 & MDF09 & \zps & 34 & 4.8 & 1.02 & 1.26 & 23.7\\
MDF04 & \yps &  8 & 4.3 & 1.03 & 1.21 & 22.0 & MDF09 & \yps & 12 & 4.3 & 0.94 & 1.12 & 22.0\\
MDF05 & \gps & 42 & 4.6 & 1.24 & 1.58 & 24.4 & MDF10 & \gps & 30 & 4.5 & 1.26 & 1.60 & 24.2\\
MDF05 & \rps & 40 & 4.6 & 1.17 & 1.46 & 24.3 & MDF10 & \rps & 33 & 4.5 & 1.18 & 1.53 & 24.2\\
MDF05 & \ips & 34 & 4.8 & 1.06 & 1.44 & 24.3 & MDF10 & \ips & 30 & 4.8 & 1.01 & 1.31 & 24.2\\
MDF05 & \zps & 27 & 4.8 & 0.99 & 1.27 & 23.6 & MDF10 & \zps & 28 & 4.8 & 1.03 & 1.24 & 23.6\\
MDF05 & \yps & 17 & 4.6 & 1.02 & 1.33 & 22.3 & MDF10 & \yps & 11 & 4.4 & 0.96 & 1.22 & 22.2\\
\end{tabular}
\label{table:depths}
\end{center}
\tablecomments{$N$ is the number of nights of observation, 
  $\log t$ is the $\log_{10}$ of the net exposure time
  in sec, ``$PSF$'' is the DoPhot FWHM of the {\it core-skirt} PSF in the
  stack-stacks, $\langle w\rangle$ is the median IPP FWHM of the observations,
  and $m_{lim}$ is the 5$\sigma$ detection limit for point sources.}

\end{table}

\section{Data processing}
\label{sec:processing}

The Pan-STARRS1 IPP system performed flatfielding on each of the
individual images, using white light flatfield images from a dome
screen, in combination with an illumination correction obtained by
rastering sources across the field of view. Bad pixel masks were
applied, and carried forward for use at the stacking stage. After
determining an initial astrometric solution, the flat-fielded images
were then warped onto the tangent plane of the sky, using a flux
conserving algorithm. The plate scale for the warped images is 0.200
arcsec/pixel. The IPP software for stacking and photometry  is still being
optimized. Therefore, for this paper we generate stacks using
customized software from one of us (JT) and produce aperture
photometry catalogs running Sextractor \citep{Bertin96} on each stacked image
independently. At a second stage we match the catalogs requiring a
detection in each band within a one arcsec radius. For simplicity we
use a rather large aperture radius (7.4 arcsec) for our photometry and
we do not apply any seeing correction, even if for some of the Medium-Deep 
fields slight variations in the FWHM between the filters are
observed (see Table \ref{table:depths}). In the production mode of
operations IPP will provide stacks with homogenized PSF, where forced
photometry will be performed at each point where a detection in one of
the unconvolved stacks is reported. The catalogs will be ingested in
the Published Science Products Subsystem \citep[PSPS,][]{Heasley08},
the database that will serve the scientific community with the final
\PS\ products. Note that at present PCS uses fluxes and flux ratios,
but no morphological information, such as spatial extent or shape.

\section{The PCS system}
\label{sec:description}

The science projects described in the Introduction define broadly the
requirements for the PCS. It should provide software tools to compute:
(a) photometric, color-based star/QSO/galaxy classification (i.e.,
morphological classifiers based on sizes and shapes are not part of
PCS, even if this additional information could and probably will be
added in the future), (b) best fitting spectral energy distribution
and photometric redshifts (photo-z) with errors for (reddish)
galaxies. Furthermore (c), (a subset of the stellar parameters)
best-fitting temperature, metallicity, gravity and interstellar
extinction with errors for (hot and cool) stars should be provided.
The codes should be interfaced to the PSPS database and to the
dataserver of IPP and results written directly into PSPS (i.e. photo-z
with errors) and into additional databases linked to the PSPS, dubbed
MYDB.

In the following, we first describe the algorithms that implement (a)
and (b) (Section \ref{sec:algorithms}), then the system components (i.e. the
different independent pieces of code that make PCS,
Section \ref{sec:components}), and finally its implementation
(Section \ref{sec:implementation}). Point (c) is still under
development and will be described in a future paper. Presently, (a)
and (b) do not communicate with each other and work independently. We
plan to upgrade the package in the future, merging in an optimal way
the classification information coming from both approaches, and using
it to improve the determination of photometric redshifts.

PCS is designed to work with catalogs providing fluxes in the \gps,
\rps, \ips, \zps, \yps\ bands and their errors (further bands can be
added if available through external datasets). Optimal performances
are expected for objects with good photometry in all \PS\ bands, but
(somewhat deteriorated) output can be obtained even in the absence of
some bands or low signal-to-noise data.

\subsection{The algorithms}
\label{sec:algorithms}

\subsubsection{SVM for PanDiSC}

There is an ongoing trend in astronomy towards larger and deeper
surveys. The natural consequence of this is the advent of very large
datasets.  There is therefore a need for automated data handling, or
data mining, techniques to handle this data volume. Automated source
classifiers based on photometric observations can provide class labels
for catalogues, or be used to recover objects for further study
according to various criteria.  

The SDSS used a
selection of algorithms to classify catalogue objects and select
followup targets \citep{2006ApJS..162...38A}.  Methods based on colour
selection were particularly employed for finding probable quasars
\citep{2002AJ....123.2945R}. More recently, \cite{gao},
\cite{richards1} and \cite{richards2} have used Kernel density
estimators for quasar selection in SDSS data.  Galaxy classification
is usually primarily based on detecting source extension by comparing
PSF magnitudes with magnitudes based on various profile models. For
examples of Galaxy classification see \cite{2011AJ....141..189V} who
used decision trees for star-galaxy separation in SDSS, or
\cite{2011MNRAS.412.2286H}, who used a Bayesian method for star-galaxy
separation in SDSS and UKIDSS. \cite{tsalmantza2007} and
\cite{tsalmantza2009} have developed a test library and selection
methods for identifying galaxies in the forthcoming Gaia mission.

\cite{Lee08} identified various stellar populations for followup
in the SEGUE survey from SDSS photometry.  Other attempts to identify stellar
populations from photometric data include \cite{2010A&A...522A..88S}, who
investigated the use of several automated classifiers on SDSS data to identify
BHB stars, and \cite{2011ApJ...726..103K} who used a support vector machine to
separate field giants from dwarfs using photometric data from a range of
surveys. \cite{2009AJ....138...63M} used a kNN technique to search for brown
dwarfs in Spitzer data. 

Finally, it should be mentioned that there is a whole industry of finding ways
to classify various types of variable objects based on their photometric light
curves.  See for example \cite{2011MNRAS.414.2602D} for classification of
various stellar types with a Random forest method, or
\cite{2010ApJ...714.1194S} who developed a method for separating quasars from
variable stars based on a structure function fit.

The \PS\ Discrete Source Classifier (PanDiSC) used here is based on a support
vector machine (SVM), a statistical learning algorithm. SVM works by
learning a nonlinear boundary to optimally separate two or more
classes of objects. Here it takes as input the 4 \PS\ \gps-\rps,
\rps-\ips, \ips-\zps, \zps-\yps\ colors. PanDiSC is based on the
Discrete Source Classifier under development for the purpose of
classifying low resolution spectroscopy from Gaia \citep{Bailer08}.
The SVM runs in the PanDiSC component of PCS (see Section
\ref{sec:components}).  The SVM implementation used is libSvm
\citep{CC11}, available at
http://www.csie.ntu.edu.tw/$\tilde{~}$cjlin/libsvm.  Probabilities are
calculated by modeling the density of data points on either side of
the decision boundary, according to the method of \citet{platt}, and
the multiclass probabilities are obtained by pairwise coupling, as
described by \citet{Wu04}. PanDiSC chooses the highest membership
probability from the eventual output for each source and assigns the
membership to the star/QSO/galaxy class accordingly.

The SVM is trained on a sample of objects with \PS\ photometry and
spectroscopic classification, to which the parameters $\gamma$ (the
scaling factor) and $C$ (the regularisation cost) of the radial basis
functions (RBF) kernel are tuned using a downhill simplex algorithm
\citep{Smith09}. The system has been applied to the SDSS DR6 dataset
\citep{Elting08} producing an excellent confusion matrix (i.e. high
accuracy of classification, $\ge96$\%, and low percentage of false
positives, $\approx 29$\% for the stellar catalog, 0.5\% for the
galaxy catalog, 10\% for the QSO catalog). 
The code is a compiled Java program.

\subsubsection{PhotoZ for PanZ}

In the last decade several efficient codes for the determination of
photometric redshifts have been developed, based either on empirical
methods, or template fitting. In the first case one tries to
parametrize the low-dimensional surface in color-redshift space that
galaxies occupy using low-order polynomials, nearest-neighbor searches
or neural networks \citep{Csabai03,Collister04}. These codes extract the
information directly from the data, given an appropriate training set
with spectroscopic information. Template fitting methods work instead with a
set of model spectra from observed galaxies and stellar population
models \citep{Padmanabhan05,Ilbert06,Mobasher07,Ilbert09,Pello09}.

The PhotoZ code used in the PanZ component of the PCS system (see
Section \ref{sec:components}) belongs to this last category and is
described in \citet{Bender01}.  The code estimates redshifts $z$ by
comparing T, a set of discrete template SEDs, to the broadband
photometry of the (redshifted) galaxies.  For each SED the full
redshift posterior probability function including priors for redshift,
absolute luminosity, and SED probability is computed using Bayes'
theorem:
\begin{equation}
\label{eq:Bayes}
P(z,T|F,M,...) \propto p(F|z,T)p(z,T|M),
\end{equation}
where $F$ is the vector of measured fluxes in different bands, $M$ the
galaxy absolute magnitude in the B band (see below), $p(z,T|M)$ the
prior distribution and $p(F|z,T)\propto exp(-\chi^2/2)$ is the
probability of obtaining a normalized $\chi^2$ for the given dataset,
redshift and template $T$. In detail, we compute $\chi^2$ as:
\begin{equation}
\label{eq:chi}
\chi^2=\sum_i\frac{(\alpha f_{i,mod}(z)-f_{i,dat})^2}{(w_i\alpha f_{i,mod})^2+\Delta f_{i,dat}^2},
\end{equation}
where $f_{i,mod}(z)$ and $f_{i,dat}$ are the fluxes of the templates (at
the redshift $z$) and of the data in a filter band $i$, and $\Delta
f_{i,dat}$ are the errors on the data. The model weights $w_i$
quantify the intrinsic uncertainties of the SEDs for the specific
filter $i$, presently they are all set to $w_i=0.1$. The normalization
parameter $\alpha$ is computed by minimizing $\chi^2$ at each choice
of parameters.
%
%
 
The priors are (products of) parameterized functions of the type:
\begin{equation}
\label{eq:Priors}
p(y)\propto y^nexp\left[-ln(2)\left( \frac{y-\hat{y}}{\sigma_y}\right)^p\right],
\end{equation}
where the variable $y$ stands for redshift or absolute magnitudes.
Typically we use $n=0$, $p=6$ or $8$, and $\hat{y}$ and $\sigma_y$
with appropriate values for mean redshifts and ranges, or mean
absolute B magnitudes and ranges, which depend on the SED type. The
absolute magnitudes of the objects are computed on the fly for the
considered rest-frame SED, normalization $\alpha$ and redshift, using
the standard cosmological parameters ($\Omega_m=0.3$, $\Lambda=0.7$,
$H_0=70$ Mpc/(km/s)).  The use of fluxes $f_{i,dat}$ instead of the
magnitudes allows us to take into account negative datapoints and
upper limits.

The set of galaxy templates is semi-empirical and can be optimized
through an interactive comparison with a spectroscopic dataset. The
original set \citep{Bender01, Gabasch04} includes 31 SEDs describing a
broad range of galaxy spectral types, from early to late to
star-bursting objects. Recently, we added a set of SEDs tailored to
fit Luminous Red Galaxies \citep[LRGs,][]{Eisenstein01}, see
\citet{Greisel11}, and one SED to represent an average QSO
  spectrum. This was obtained by averaging the low redshift HST
  composite of \citet{Telfer02} and the SDSS median quasar composite
  of \citet{Vandenberk01}. Furthermore, the method also fits a set of
stellar templates, allowing a star/galaxy classification and an
estimate of the line-of-sight extinction for stellar objects. The
templates cover typically the wavelength range $\lambda=900$ \AA\ up
to 25000 \AA\ (with the QSO template covering instead 300-8000 \AA)
and are sampled with a step typically 10 \AA\ wide (varying from 5 to
20 \AA; the QSO SED has $\Delta \lambda=1$ \AA).

The method has been extensively tested and applied to several
photometric catalogs with spectroscopic follow-up \citep{Gabasch04,
  Feulner05, Gabasch08, Brimioulle11}.  Given a (deep) photometric
dataset covering the wavelength range from the U to the K band,
excellent photometric redshifts with $(z_{phot}-z_{spec})/(1+z)\sim0.03$ up to
$z\approx 5$ with at most a few percent catastrophic failures can be
derived for every SED type. With the help of appropriate priors,
photometric redshifts accurate to 2\% (in
$(z_{phot}-z_{spec})/(1+z_{spec})$) with just 1\% outliers (see
Sect. \ref{sec:results} for definitions) are obtained for LRGs  using
the ugriz SDSS photometric dataset \citep{Greisel11}. This is the
first time we attempt to determine the photometric redshifts of QSOs.
We couple the available SED to a strong prior in luminosity that dampens
its probability as soon as the predicted B band absolute magnitude is
fainter than -24.

The code is in C++. A Fortran version is available as implemented
  under Astro-WISE \citep{Valentijn06,Saglia11}.



\subsection{The components of PCS}
\label{sec:components}

\begin{figure}[htbp]
\begin{center}
\centerline{\includegraphics[width=22cm,angle=90]{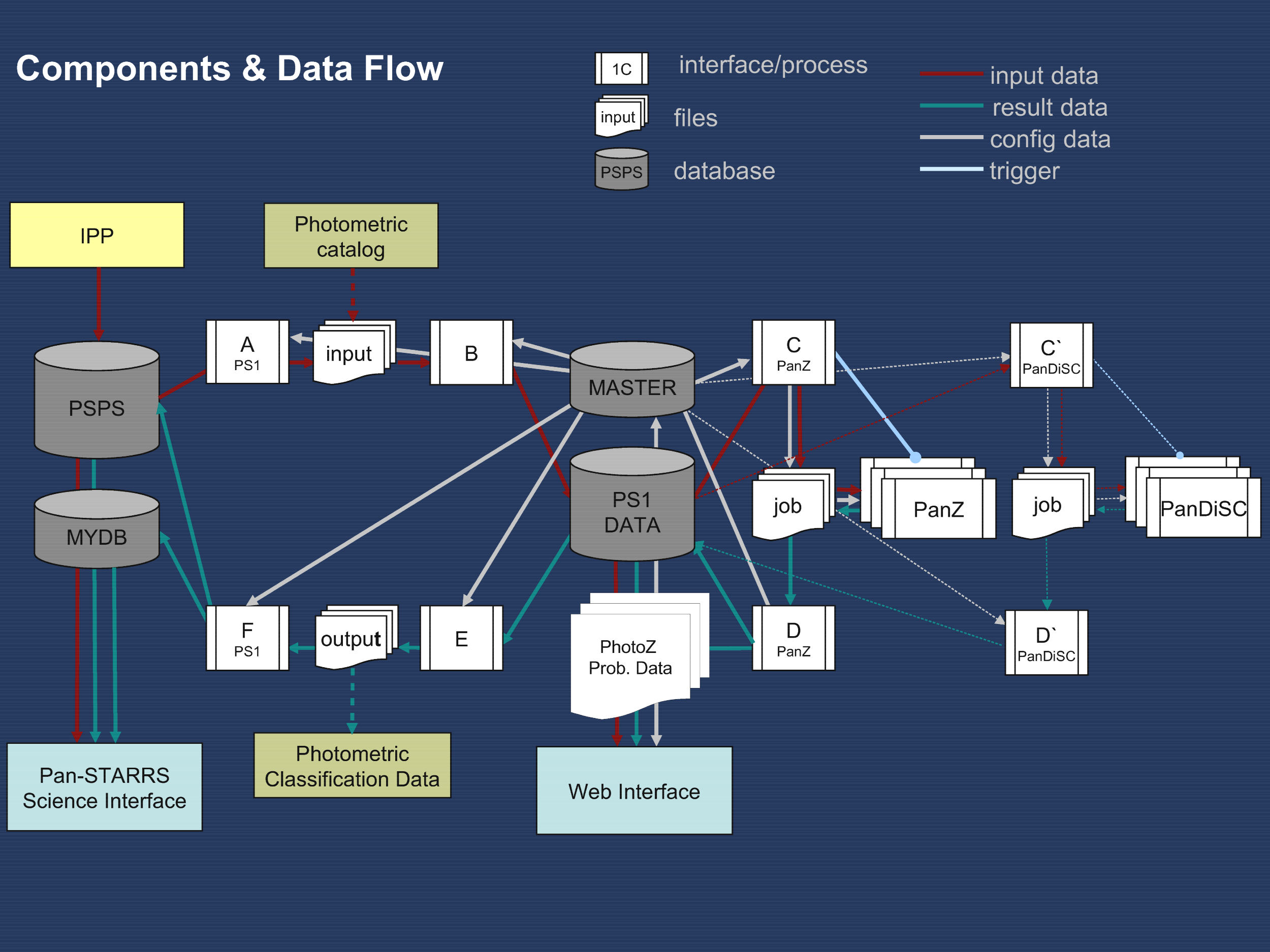}}
\caption{The components of the PCS, see Sect. \ref{sec:components} 
for a description.}
\label{fig:PCScomponents}
\end{center}
\end{figure}

Fig. \ref{fig:PCScomponents} describes schematically the components 
and data flow of
PCS. Each {\it component}, or module, is a separate
unit of the package with a well defined operational goal. Following
the usual convention, it is indicated by a box with bars (for
interfaces and/or processes, from A to F, plus PanZ and PanDiSC) or a
cylinder (for a database) in Fig. \ref{fig:PCScomponents}. There are
four databases: two reside in Hawaii (PSPS, where the primary \PS\
catalogs are stored and a subset of the output produced by PCS is
copied, and MYDB, where the whole of the PCS output goes). The other
two are in Garching: the Master, where configuration files and
templates are stored, and the local PS1 DATA, where PCS input and output
are stored. Light-blue boxes indicate
interfaces to the users and the yellow box to the upper left the IPP
system.  The {\it arrows} in the figure represent links between the
components. Their colors code the type of link (red for input, blue
for output/results, grey for configuration data, cyan for a
trigger). For clarity the lines joining to the C$'$, D$'$ and PanDiSC
components are dotted. The paper-like symbols indicate generated
data-files. The two yellow boxes ``Photometric catalog'' and
``Photometric Classification Data'' refer to the manual mode of
operations, see below.

In the normal batch mode of activities, the interfaces/processes A to F
permanently run in the background and react to changes in the PSPS
database (A) or the Master database in stand-alone mode.  However,
parallel manual sessions can be activated, where the user is free to
use parts or all of the pipeline, adding further photometric or
spectroscopic datasets to the local database, changing setups, testing
new SEDs and recipes or defining and using new training sets. Here below
we first describe the automatic mode of operations and then summarize
the manual options.

The starting point is the PSPS database in Hawaii, which is filled
with catalog data produced by IPP. The module A periodically checks
when new sets of data with all 5 band fluxes measured are available in
PSPS. It copies tables of input data according to selectable input
parameters to Garching. The module B detects the output of process A,
ingests these data in the local database and computes for each object
the galactic absorption corrections according to the
\citet{Schlegel98} maps. They are applied to the photometry only when
computing photometric redshifts. The bare photometry is considered
when classifying the objects (in PanDiSC) or when testing the stellar
templates (in PanZ). Once B has finished, the module C and C' start to
prepare PanZ and PanDiSC jobs, respectively, to analyze the newly
available data, according to the preset configuration files. The input
catalog is split into many suitable chunks to allow the triggered
submission of multiple jobs on the parallel queue of the computer
cluster (see Sect. \ref{sec:implementation}).  Once all the jobs are
finished, the modules D and D$'$ become active. They copy the results
of PanZ and PanDiSC, respectively, into the local database, excepting
the full redshift probability distributions for each object, that are
too large to be written in the database directly. They are saved as
separate compressed files. At this point the procedure E is
activated. This prepares the data tables for delivery to
Hawaii. Finally the module F signals PSPS in Hawaii, which then
uploads them into the PSC MYDB part of the PSPS database. They can be
accessed by the whole \PS\ community through the \PS\ Science
Interface (the light blue box at the lower left of
Fig. \ref{fig:PCScomponents}).

As a second mode of operation, the system can work in 'manual' mode,
where specific data sets can be (re-)analyzed independently of the
automatic flow. This is useful in the testing phase, or while
considering external datasets, such as the SDSS, or mock catalogs of
galaxies (indicated by the yellow box ``Photometric catalog'' of
Fig.\ref{fig:PCScomponents}). The output of such manual runs is
indicated by the yellow box ``Photometric Classification Data'' of
Fig.\ref{fig:PCScomponents}. Many of the tests discussed in this paper
have been obtained in 'manual' mode. Extensive spectroscopic datasets
are available in the local database to allow an efficient
cross-correlation with the \PS\ objects.

The configuration files, as well the different modes of operations,
can be manipulated through a user-friendly web interface (the light
blue box at the lower right of Fig. \ref{fig:PCScomponents}). This
feature, initially developed to ease the life of the current small
number of PCS members, makes the system interesting for a possible future
public release.

\subsection{The implementation of PCS}
\label{sec:implementation}

The PCS is implemented on the PanSTARRS cluster, a 175 node (each with
2.6GHz, 4 CPUs and 6 GB memory, for a total of 700 CPUs) Beowulf
machine with 180 TB disk space, attached to a PB robotic storing
device, mounted at the Max-Planck Rechenzentrum in Garching.  The
modules described in the previous section are a series of shell
scripts, or html/php files, executing php code or SQL commands, or
running compiled C++ code. In particular, the input/output interfaces
A and F to the PSPS database make use of SOAP/http calls. The Schlegel
maps are queried using the routines available from the web. The local
database is based on MySQL. A set of Python scripts allows the user to
automatically generate plots and statistics similar to
Fig. \ref{fig:PanZSLOANFields} and \ref{fig:PanZSLOAN}. Presently, new
available photometry is downloaded from the PSPS database in chunks of
4 million objects. They are split in blocks of 25000, each of which is
sent as a single job to the parallel queue. These numbers are subject
to further optimization. The system allows parallel running of
jobs operating on the same dataset, but with different recipes.

The performance (in terms of processed objects per second) of the
complete system is summarized in the Table \ref{table:performances},
where the case of a catalog of 1.8 million objects is presented.

The modular structure of the PCS allows implementation of
further components, such as alternative photometric redshift codes and the
modules to constrain stellar parameters that are under development.

\begin{table}[htdp]
\caption{Performance test for the PCS based on a catalog with 1.8 million 
objects. }
\begin{center}
\begin{tabular}{llrr}
\hline
\hline
Module & Task                       & Duration & Performance \\
\hline
A     & Read from PSPS             & 4 min    & 7500/sec\\
B     & File conversion            & 2:10 min & 14000/sec\\
B     & Input data injection       & 5:20 min & 5600/sec \\
C     & Extraction (18 data files) & 2:10 min & 14000/sec \\
C     & Job creation (18 jobs)     & 0:30 min & -\\
PanZ  & Running jobs               & 32 min   & 950/sec \\
D     & Results injection          & 16 min   & 1875/sec \\
D     & PhotoZ-P compression + storage & 75 min & 400/sec\\
C$'$  & Extraction (18 data files) & 2:00 min &  15000/sec\\
C$'$  & Job creation (18 jobs)     & 0:30 min & -\\
PanDiSC  & Running jobs            & 1:00min   & 30000/sec \\
D$'$  & Results injection          & 15 min   &  2000/sec\\
E     & Results extraction         & 1:40 min & 18000/sec \\
F     & Signal to MyDB        & -         & -\\
F     & Signal to PSPS        & -         & -\\
\hline
\end{tabular}
\end{center}
\label{table:performances}
\end{table}

\section{PCS tests}
\label{sec:results}

Several spectroscopic surveys overlap with the \PS\ MDFs, providing
abundant spectral classifications and redshifts: BOSS \citep[MDF1, 4,
9, 10]{Aihara11}, CDFS \citep[MDF2]{Vanzella06}, SDSS and SEGUE/SDSS
\citep[MDF2 to 10]{Eisenstein01, Lee08, SDSS}, 2dF Galaxy Redshift
Survey \citep[MDF4]{Colless01}, VVDS \citep[MDF1, 7 and
10]{Lefevre04,Lefevre05,Garilli08}, ZCosmos \citep[MDF4]{Lilly07}.
Here we present tests of PanDiSC and PanZ based on the DR7 SDSS dataset, 
that provides the largest available homogeneous spectroscopic follow-up of LRGs
(selected as DR7 objects with primTarget=32 or 96),
the objects for which the \PS\ filter set suffices to deliver
excellent photometric redshifts. Moreover, the SDSS dataset comprises
a large enough sample of stars and QSOs to allow a global sensible
test of the capabilities of PCS. The remaining spectroscopic surveys
will be discussed in future \PS\ papers dealing with the science
applications of the PCS.

\begin{figure}[htbp]
\begin{center}
\centerline{
\includegraphics[width=16cm]{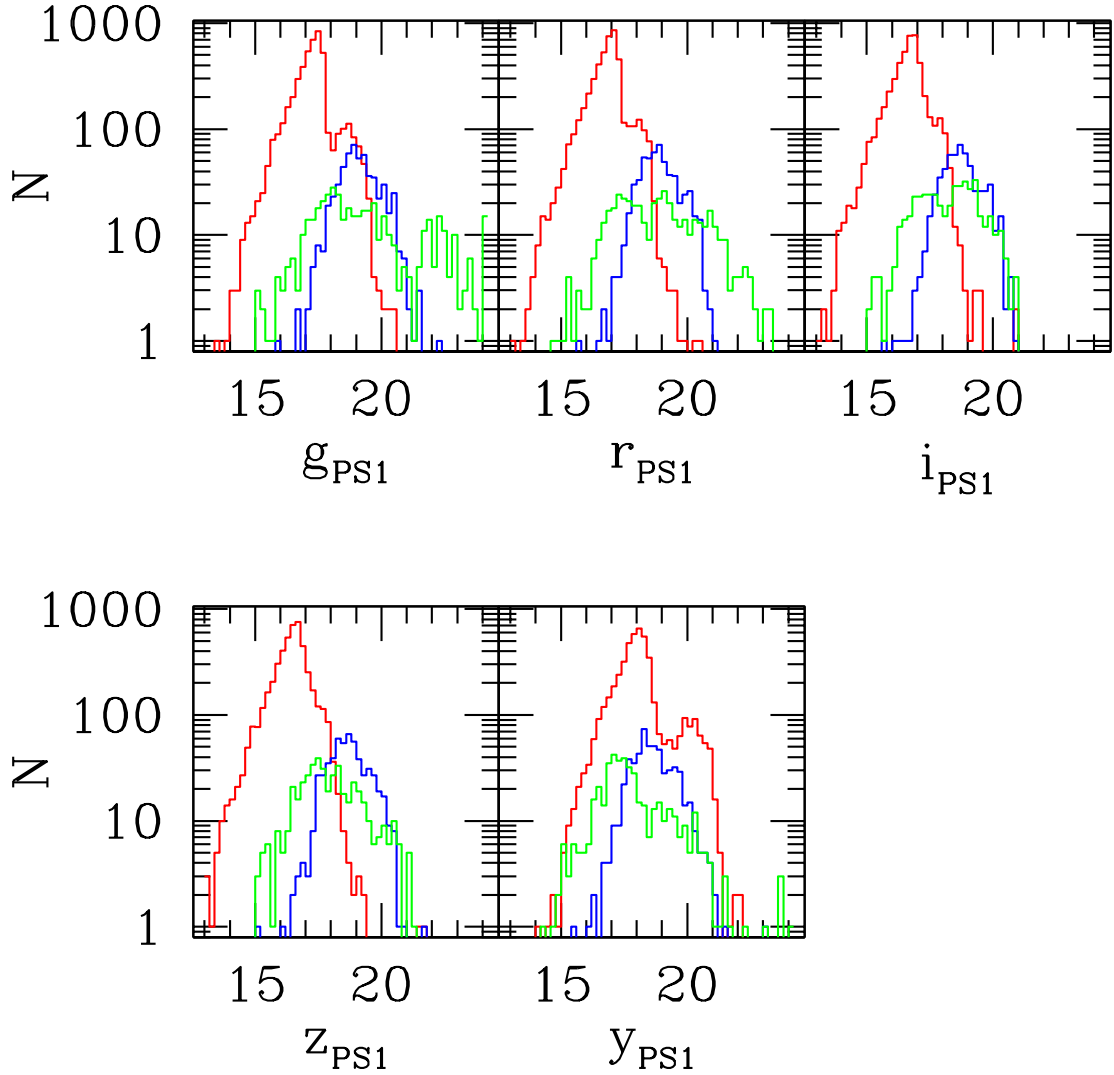}
}
\caption{The \PS\ magnitude (inside an aperture of 7.4 arcsec radius)
  histograms matched to the spectroscopic SDSS and classified through
  the spectroscopic SDSS information as stars (green), galaxies (red)
  and QSOs (blue). }
\label{fig:mag}
\end{center}
\end{figure}

Table \ref{tab:spectra} gives the average galactic reddening of the
fields covered by SDSS (the largest value of $E(B-V)=0.066$ mag is
reached in MDF09) and summarizes the numbers of matched objects,
typically several hundreds per field, with more than a thousand in
MDF4 and a total of 5784. The same table splits them into stars,
galaxies and QSOs.  The majority of the matches are galaxies (of which
approximately a quarter are red), but numerous (of the order of a
hundred) stars and quasars are represented per field.
Fig. \ref{fig:mag} shows the histograms of the magnitudes in the \PS\
filters (within an aperture of 7.4 arcsec radius) for the SDSS
spectroscopically classified (and assumed to be the ``truth'') stars,
galaxies and QSOs of the sample. As expected due to the SDSS
spectroscopic limits (resulting from the main spectroscopic sample
limited at r=17.77 and the sparser additional surveys, reaching
fainter magnitudes), the galaxy sample peaks around \gps$\approx
17.5$, while the QSO sample is $\approx 1.5$ mag fainter. The star
sample spans a broader range of magnitudes, from \gps$\approx 16$ to
$\approx 24$. Given the achieved depth of the MDFs, photometric data
for this sample has very large signal-to-noise and can be used to test
the systematic limitations of PCS. In the following we describe tests
performed using both the \PS\ and the SDSS photometry, to show that
the \PS\ dataset is at least as good as SDSS.

\begin{table}[htdp]
\caption{The SDSS spectroscopic dataset used to test PCS, with number of objects split by star/galaxy/QSO category (LRGs are a subset of galaxies).}
\begin{center}
\begin{tabular}{rcrrrrr}
\hline
\hline
MDF   & $\langle E(B-V)\rangle$ & $N_{SDSS}$ &  Stars & Galaxies & LRGs & QSOs \\  
      & (mag)                   &           &        &          &      & \\
\hline                               \hline                                     
3     & 0.027 & 704   & 49    &  577     & 107  & 78   \\  
4     & 0.026 & 1125  & 128   &  880     & 169  & 117   \\ 
5     & 0.008 & 913   & 41    &  785     & 155  & 87   \\  
6     & 0.014 & 732   & 27    &  628     & 153  & 77   \\  
7     & 0.011 & 953   & 104   &  755     & 173  & 94   \\  
8     & 0.010  & 226   & 2     &  207     & 49   & 17   \\    
9     & 0.066 & 589   & 54    &  495     & 95   & 40   \\    
10    & 0.038 & 542   & 51    &  436     & 99   & 55   \\  
Total &       & 5784  & 456   & 4763     & 1000 & 565   \\
\hline
\end{tabular}
\end{center}
\label{tab:spectra}
\end{table}

\subsection{Star/Galaxy/Quasar classification}
\label{sec:testPanDiSC}

We trained and tested the PanDiSC SVM using a 10-fold
  cross-validation approach. We divided the sample described in Table
  \ref{tab:spectra} with \PS\ photometry in 10 partitions, each with
  45 stars, 475 galaxies and 55 QSOs. Note that they span the
  magnitude ranges shown in Fig. \ref{fig:mag}.  We constructed 10
  subsets by randomly selecting 55 out of the 475 galaxies. We created
  10 training sets, by concatenating 9 of the subsets, and tested the
  PanDiSC SVM on each of the 10 partitions not used in the 10 training
  sets.

Table \ref{table:confusionmatrix} shows the results: galaxies are
  classified correctly in almost 97\% of cases, with very small
  variations from field to field. Stars (one of which having equal
  probability of being a star, a galaxy or a QSO, and therefore not
  considered) and QSOs are recovered on average in $\approx 84$ \% of
  cases.  Successful star classification goes up to 95\% for MDF5, and
  is as low as 75\% for MDF10.  Successful QSO classification is at
  the 87\% level in MDF4 and drops to 69\% for MDF9. We will
  investigate the statistically significance and possible cause of
  these variations with future larger \PS\ catalogs.

The relatively poor performance of the stellar
classification is driven by early-type stars alone (correctly
classified in 80\% of cases) and the \PS\ filter set.  The
classification of late-type stars is better, with just 4\% of the late
stars being wrongly classified as galaxies.

We now look at the fraction of false positives. Only  21 stars and 44
QSOs are classified as galaxies, therefore the galaxy sample defined
by PanDiSC is contaminated at the level of 1\%. This is not unexpected,
because the extragalactic MDF number counts are dominated by
galaxies. The purity of the star and QSOs samples are much worse. There
are 38 galaxies and 47 QSOs classified as stars, which results in a
19\% contamination of the star catalog defined by PanDiSC. Without
the contribution of galaxies, which could be flagged by adding
morphological information (i.e. whether the objects are point-like or
extended), the contamination (by QSOs) reduces to 10\%. Preliminary
tests \citep{Klement09} show that this development
is indeed very promising. There are 47 stars and 107 galaxies
classified as QSOs. This means that the QSO catalog defined
by PanDiSC is contaminated at the 28\% level, without galaxies just
8.5\%. Clearly, the situation will be different at lower galactic
latitudes, where stars dominate the number counts at these magnitude
limits. Overall, the results are not much worse than those reported by
\citet{Elting08}. They can be improved when larger \PS\ catalogs
will be available, by optimizing the probability thresholds for making
a classification decision, since they depend on the relative
distribution of stars, galaxies, quasars in the training/test data
sets used to make the assessment.

\begin{figure}[htbp]
\begin{center}
\centerline{
\includegraphics[width=16cm]{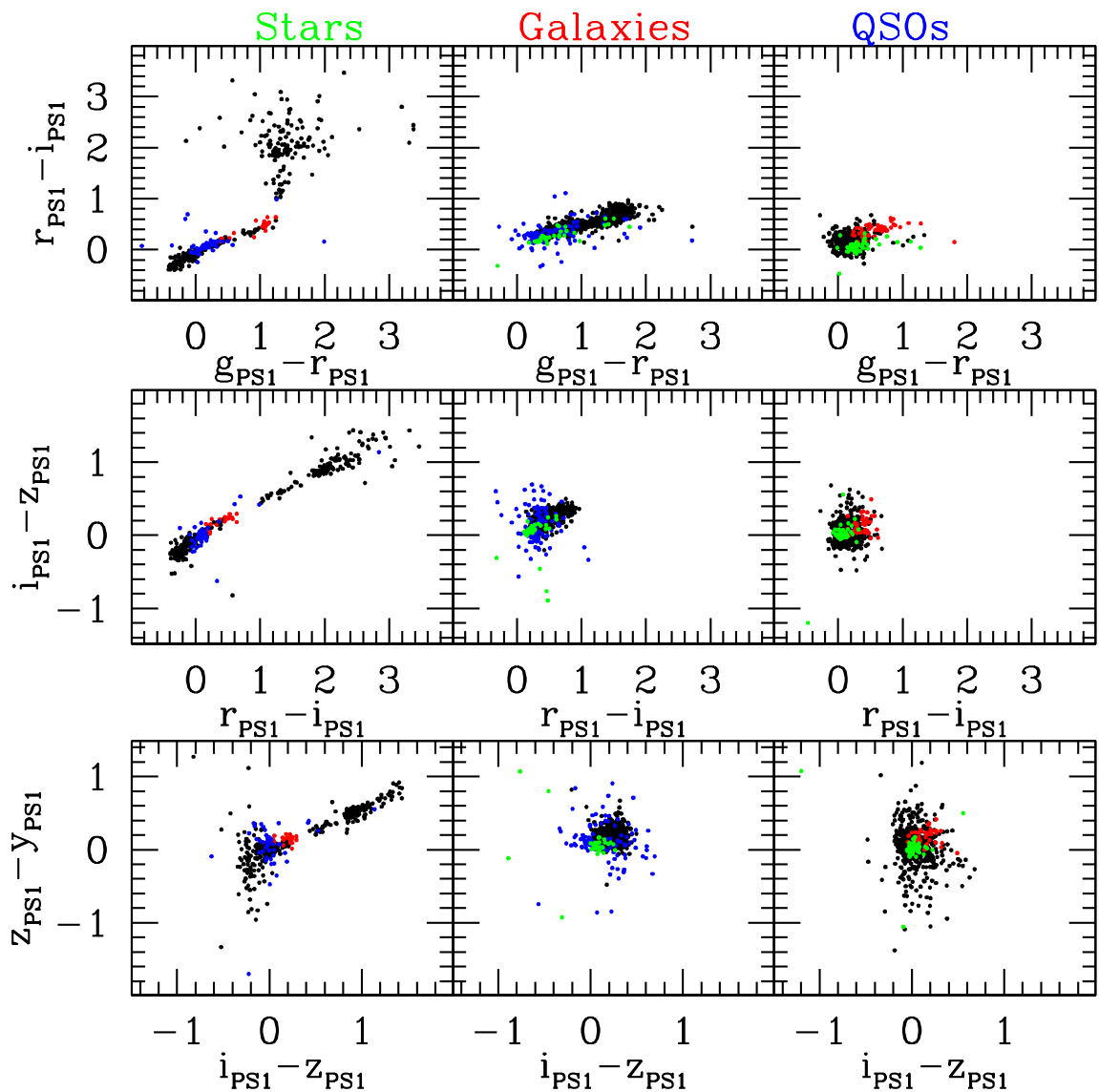}
}
\caption{The distribution of SDSS stars (left), galaxies (middle) and
  QSOs (right) in the color-color plots of \PS. Objects classified correctly 
  by PanDiSC are shown black, objects classified wrongly as stars are green,
  wrongly as galaxies are red, wrongly as QSO are blue. 
}
\label{fig:colcol}
\end{center}
\end{figure}

Fig. \ref{fig:colcol} presents the \PS\ color-color plots with the
distributions of stars, galaxies and quasars of the spectroscopic SDSS
dataset. Objects that are stars according to the spectroscopic SDSS
classification are shown on the left diagrams, objects that are
galaxies in the central diagrams and objects that are QSOs on the
right diagrams. Objects classified correctly by PanDiSC are shown
black. Objects wrongly classified as stars by PanDiSC are shown green,
wrongly classified as galaxies red, wrongly classified as QSOs in
blue.  So a red dot on the left column is a star that PanDiSC
classified as a galaxy.

Clearly, misclassifications happen in regions of color space where the
different types overlap, where only an additional filter (for example
the u band) would help discriminate between the populations. In
contrast, late-type stars are seldom misclassified thanks to their red
colors that divide them well from galaxies and QSOs. Note that the
increased scatter in the stellar \gps-\rps colors at \rps-\ips$>1.8$
is due to stars near or below the \gps\ and \rps\ magnitude limits of
Table \ref{table:depths}.

We repeated the same tests based on the 10-fold cross-validation
  technique using the SDSS Petrosian ugriz photometry: the results are
  presented in Table \ref{table:confusionmatrixSLOAN}. While the
  presence of the u band boosts the success rate for QSOs (up to 94\%)
  and stars (up to 92\%, with early type stars correctly classified in 90\% of the cases) the absence of the y band and the lower
  quality of the z band penalizes marginally the galaxy classification
  (down to 95\%) and the stellar classification of late star types
  (classified correctly in 95\% of cases). The star false
  positives are up to 28\% (of the total true stars), due to the
  higher number of misclassified galaxies, but would drop to just the
  6\% of misclassified QSOs if information about size were added. The
  QSO false positives, slightly better at the 27\%, 
  drop to just 3\% without galaxies. The galaxy false positives stay at
  the 1\% level.

\begin{table}[htdp]
  \caption{SVM predictions for evaluation set using \PS\ photometry: 
the confusion matrix (first in absolute numbers, and second in fractions
normalized to 1) with true classes in rows. }
\begin{center}
\begin{tabular}{lrrrr}
\hline
\hline
True classes & $N_{tot}$ & Star  & Galaxy & Quasar \\
\hline
Star         & 449     & 381   & 21    & 47     \\
             &         & 0.849 & 0.047 & 0.104  \\
Galaxy       & 4750    & 38    & 4605  & 107     \\
             &         & 0.008 & 0.970 & 0.022  \\
Quasar       & 550     & 47    & 44    & 459    \\
             &         & 0.085 & 0.080 & 0.835  \\
\hline
\end{tabular}
\end{center}
\label{table:confusionmatrix}
\end{table}

\begin{table}[htdp]
  \caption{SVM predictions for evaluation set using the SDSS Petrosian 
    magnitude ugriz photometry: the confusion matrix  (first in absolute 
numbers, and second in fractions
    normalized to 1) is given as fractions 
    normalized to 1 with true classes in rows. }
\begin{center}
\begin{tabular}{lrrrr}
\hline
\hline
True classes & $N_{tot}$ & Star & Galaxy & Quasar \\
\hline
Star         & 450      & 412   & 23    & 15 \\
             &          & 0.916 & 0.051 & 0.033  \\
Galaxy       & 4750     & 99    & 4525  & 126 \\
             &          & 0.021 & 0.953 & 0.026  \\
Quasar       & 550      & 25    & 15    & 510 \\
             &          & 0.046 & 0.027 & 0.927  \\
\hline
\end{tabular}
\end{center}
\label{table:confusionmatrixSLOAN}
\end{table}

\subsection{Photometric redshifts}
\label{sec:PanZ}

The \PS\ photometric dataset misses the u band on the blue side of the
spectral energy distribution, and the NIR colors redder than the y
band. Therefore one expects any photometric redshift program to
perform best for red galaxies at moderate redshifts (e.g., the LRGs), 
and to fail especially when blue galaxies at
low redshifts are considered. Similarly, late stars are expected to be
better recognized than earlier ones. Finally, since at present the
PhotoZ program misses SEDs optimized for QSOs, we expect poor
performances in these cases.

\begin{figure}[htbp]
\begin{center}
\centerline{
\includegraphics[width=16cm]{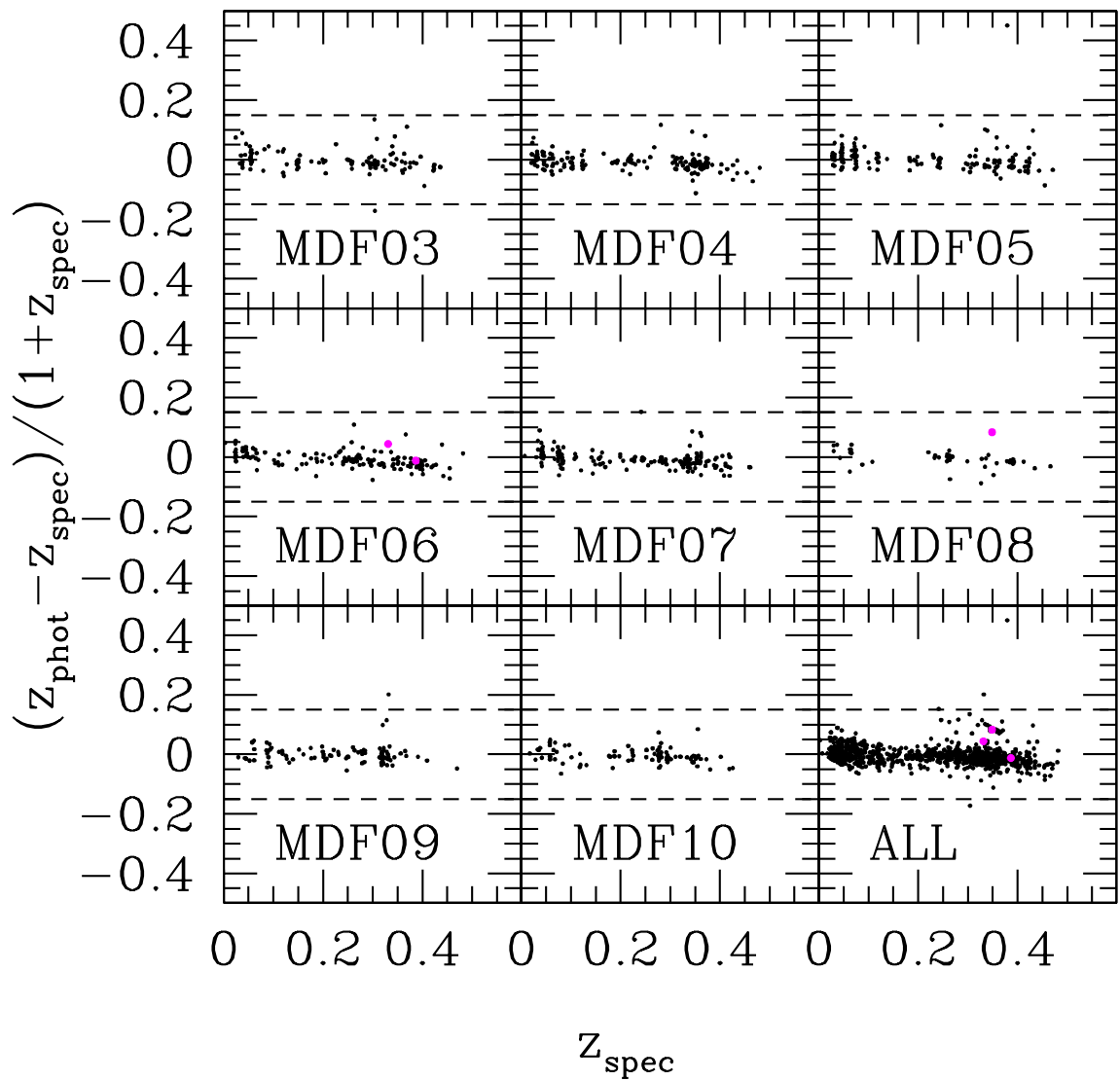}
}
\caption{The comparison between spectroscopic and photometric
  redshifts for the SDSS sample of LRGs. The difference
  $(z_{phot}-z_{spec})/(1+z_{spec})$ is shown as a function of
  $z_{spec}$. The magenta points show the three objects for which the QSO
  SEDs would give the best fit (but a catastrophically poor
  photometric redshift).}
\label{fig:PanZSLOANFields}
\end{center}
\end{figure}

\begin{figure}[htbp]
\begin{center}
\centerline{
\includegraphics[width=8cm]{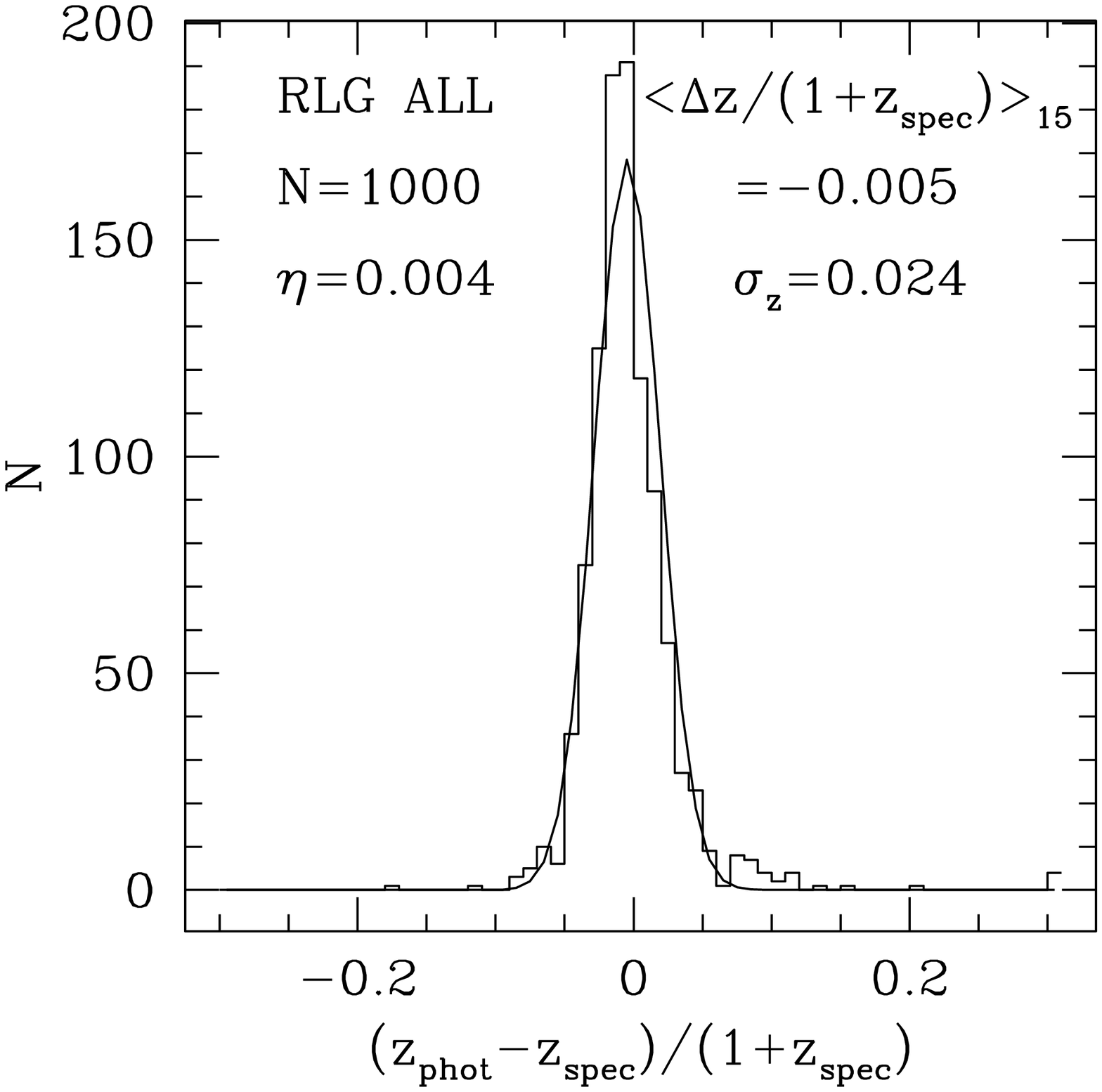}
\includegraphics[width=8cm]{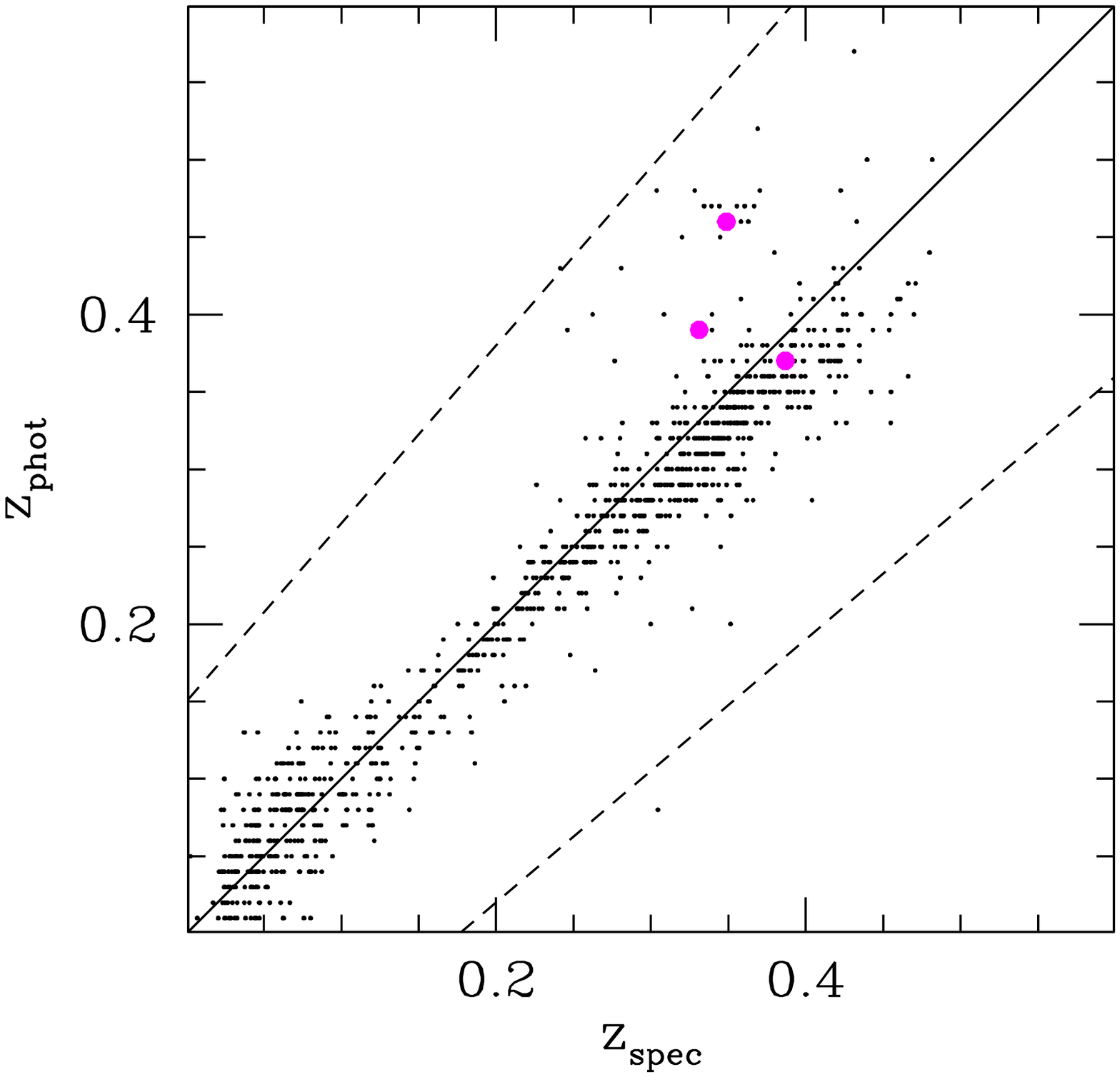}
}
\caption{The comparison between spectroscopic and photometric
  redshifts for the SDSS sample of LRGs. Left: the histogram of
  $z_{phot}-z_{spec}/(1+z_{spec})$. Right: $z_{spec}$
  vs. $z_{phot}$. The magenta points show the three objects for which the
  QSO SEDs would give the best fit (but a catastrophically poor
  photometric redshift).}
\label{fig:PanZSLOAN}
\end{center}
\end{figure}

Fig. \ref{fig:PanZSLOANFields} shows the comparison between
spectroscopic and photometric redshifts obtained for the SDSS sample
of LRGs, field by field.  Fig. \ref{fig:PanZSLOAN} shows the results
for the whole sample. As usual, we define the percentage of
catastrophic failures $\eta$ as the fraction of objects (outliers) for which
$|z_{phot}-z_{spec}|>0.15\times(1+z_{spec})$, the residual bias as the
mean of $(z_{phot}-z_{spec})/(1+z_{spec})$ without the outliers, and
the robust error as $\sigma_z=1.48\times
Median|z_{phot}-z_{spec}|/(1+z_{spec})$ without the outliers.  For
LRGs, the \PS\ photometry allows the determination of photometric
redshifts accurate to 2.4\% in $\sigma_z$, with bias smaller than 0.5\%,
no strong trend with redshift and 0.4\% catastrophic outliers (when no
QSO SED is allowed). No field-to-field dependencies are
present.

\begin{figure}[htbp]
\begin{center}
\centerline{
\includegraphics[width=8cm]{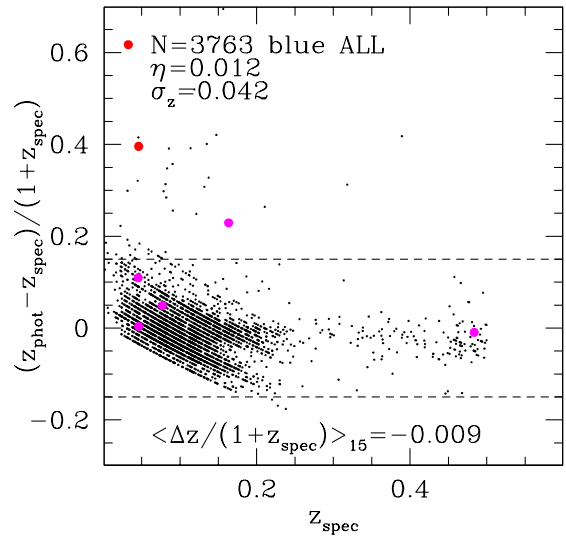}
}
\caption{The comparison between spectroscopic and photometric
  redshifts for the SDSS galaxy sample not classified as LRG.  
The magenta points show the five of the six objects for which the QSO
  SEDs would give the best fit (but a catastrophically poor
  photometric redshift, two of these cases are visible in red).}
\label{fig:PanZSLOANBlue}
\end{center}
\end{figure}

As expected, the situation is not as satisfactory for
non-LRGs. Fig. \ref{fig:PanZSLOANBlue} shows that especially at
$z_{spec}<0.2$ the residuals are biased in a systematic way and more
than 1\% catastrophic outliers are present. Nevertheless, the robust
estimate of the scatter remains below 5\%.

If we now use for the same galaxies the Petrosian ugriz SDSS
photometry, we find the following. The photometric redshifts for LRGs
are similarly good (2.6\%), but with a higher percentage of
outliers. In contrast, precision (3.7\%) and percentage of outliers
(1\%) are better for blue galaxies, where the presence of the u band
helps.

\begin{figure}[htbp]
\begin{center}
\centerline{
\includegraphics[width=8cm]{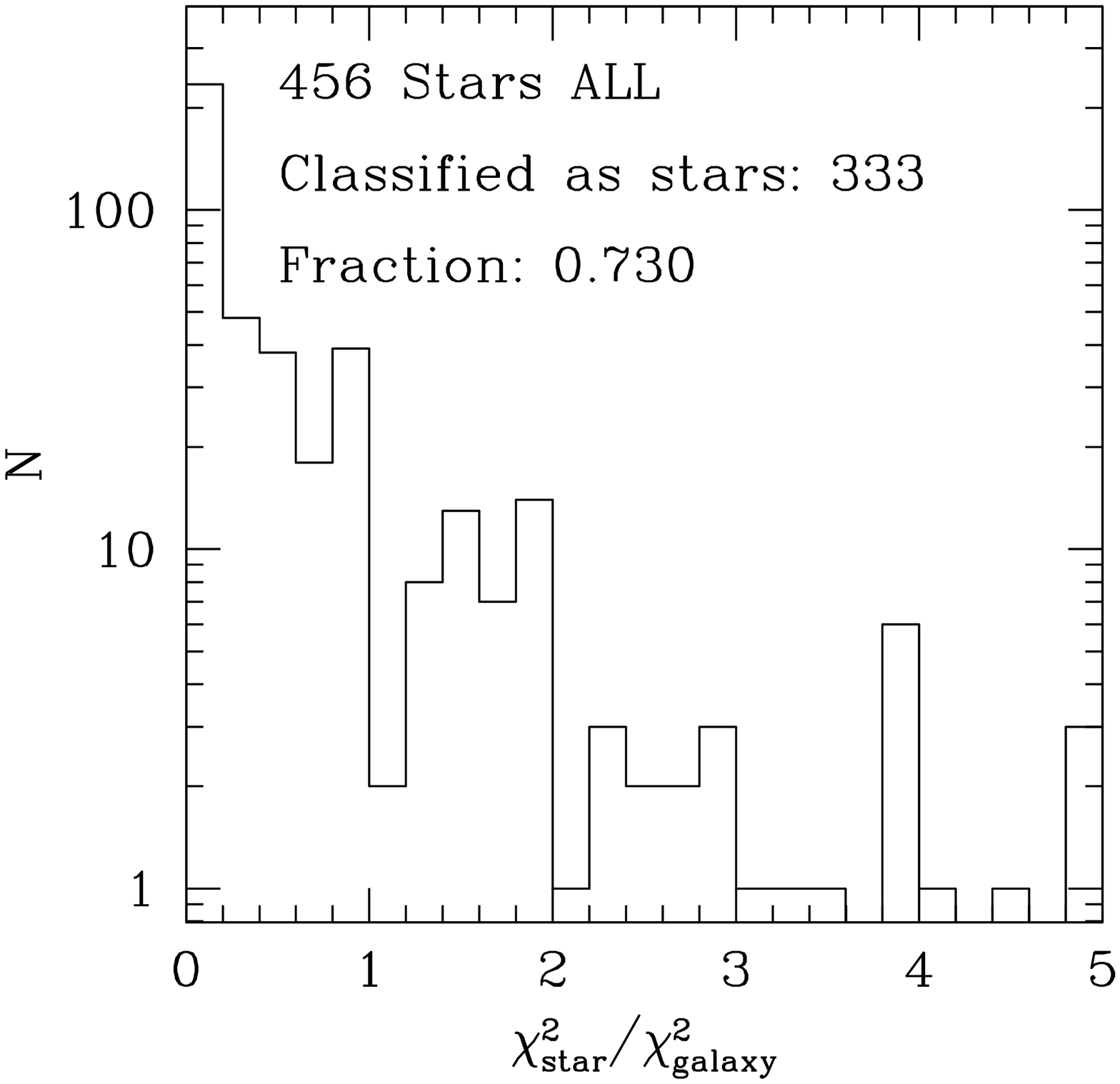}
\includegraphics[width=8cm]{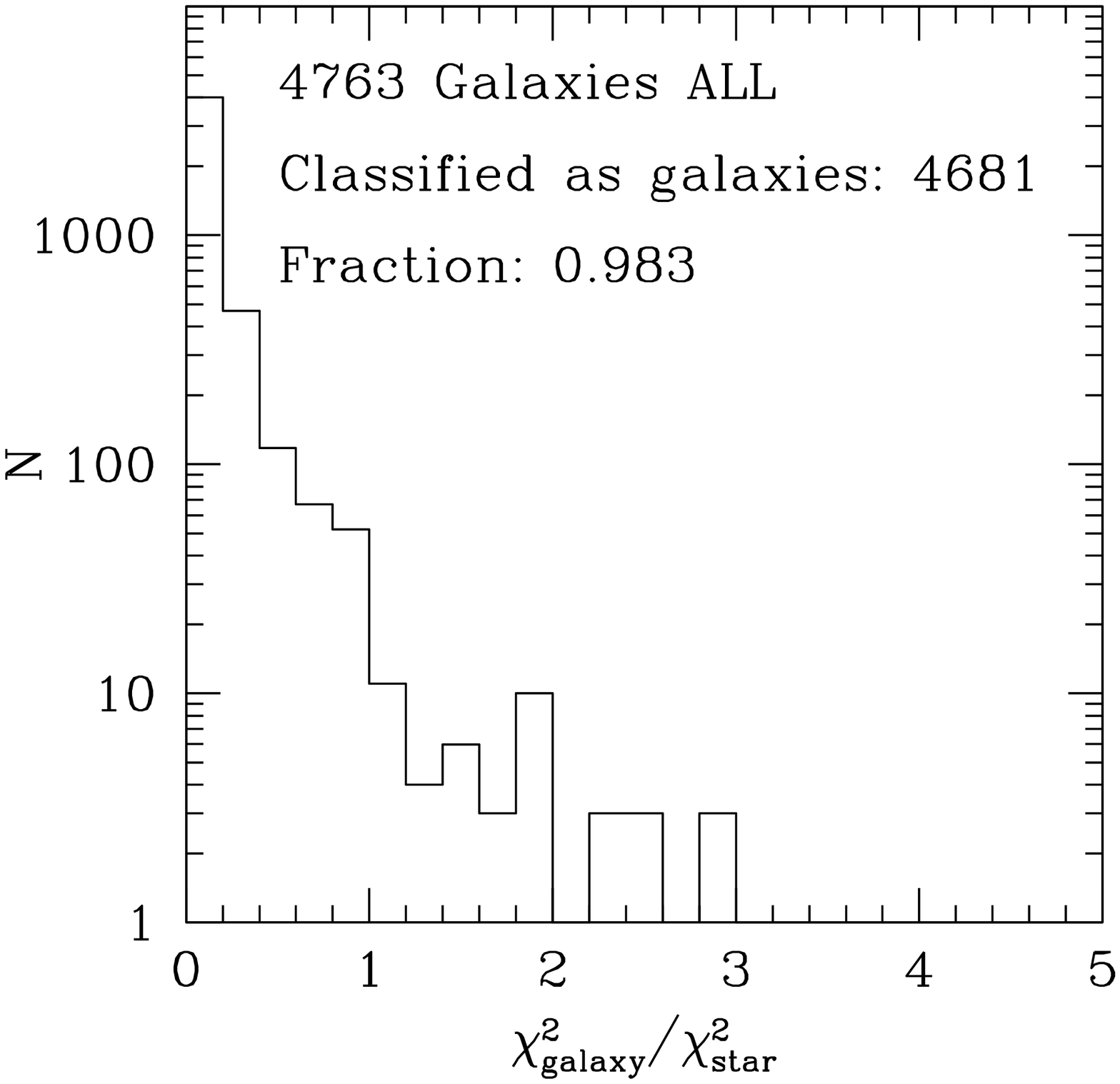}
}
\caption{The efficiency of the PanZ star recognition. Left: for 73\%
  of spectroscopically confirmed SDSS stars PanZ finds a stellar SED
  as the best fit to the \PS\ photometry
  (i.e. $\chi^2_{star}<\chi^2_{galaxy}$, note that for plotting
  convenience $\chi^2$ ratios are shown). Of the remaining 123 stars,
  13 are best-fit by the QSO SED.  Right: PanZ finds
  $\chi^2_{galaxy}<\chi^2_{star}$ for 98\% of spectroscopically
  confirmed SDSS galaxies. For 8 of these the QSO SED fits best. For
  one galaxy the best extragalactic fit is obtained by the QSO SED and
  is poorer than the one obtained using stellar templates
  (i.e. $\chi^2_{galaxy}>\chi^2_{star}$). }
\label{fig:PanZSLOANStars}
\end{center}
\end{figure}

As described in Sect. \ref{sec:algorithms}, PanZ computes also the
goodness of fits for a number of stellar templates. Therefore the
difference $\chi^2_{star}-\chi^2_{galaxy}$ between the $\chi^2$ of the
best-fitting stellar SED $\chi^2_{star}$ and the best fitting galaxy
SED $\chi^2_{galaxy}$ provides a crude galaxy/star classification: if
$\chi^2_{star}-\chi^2_{galaxy}<0$, the stellar template is providing a
better fit than the galaxy ones and we classify the object as a star.
In Fig. \ref{fig:PanZSLOANStars}, left (where just for plotting
convenience we give $\chi^2_{star}/\chi^2_{galaxy}$) we show that
requiring $\chi^2_{star}-\chi^2_{galaxy}<0$ (i.e.
$\chi^2_{star}/\chi^2_{galaxy}<1 $ in the plot) allows to correctly
classify spectroscopically confirmed SDSS stars in 73\% of the cases.
The percentage of successful classifications grows to 89\% if only
late type SDSS stars are considered. The percentage of success is 98\%
if spectroscopically confirmed SDSS galaxies are considered (Fig.
\ref{fig:PanZSLOANStars}, right).

\begin{figure}[htbp]
\begin{center}
\centerline{
\includegraphics[width=8cm]{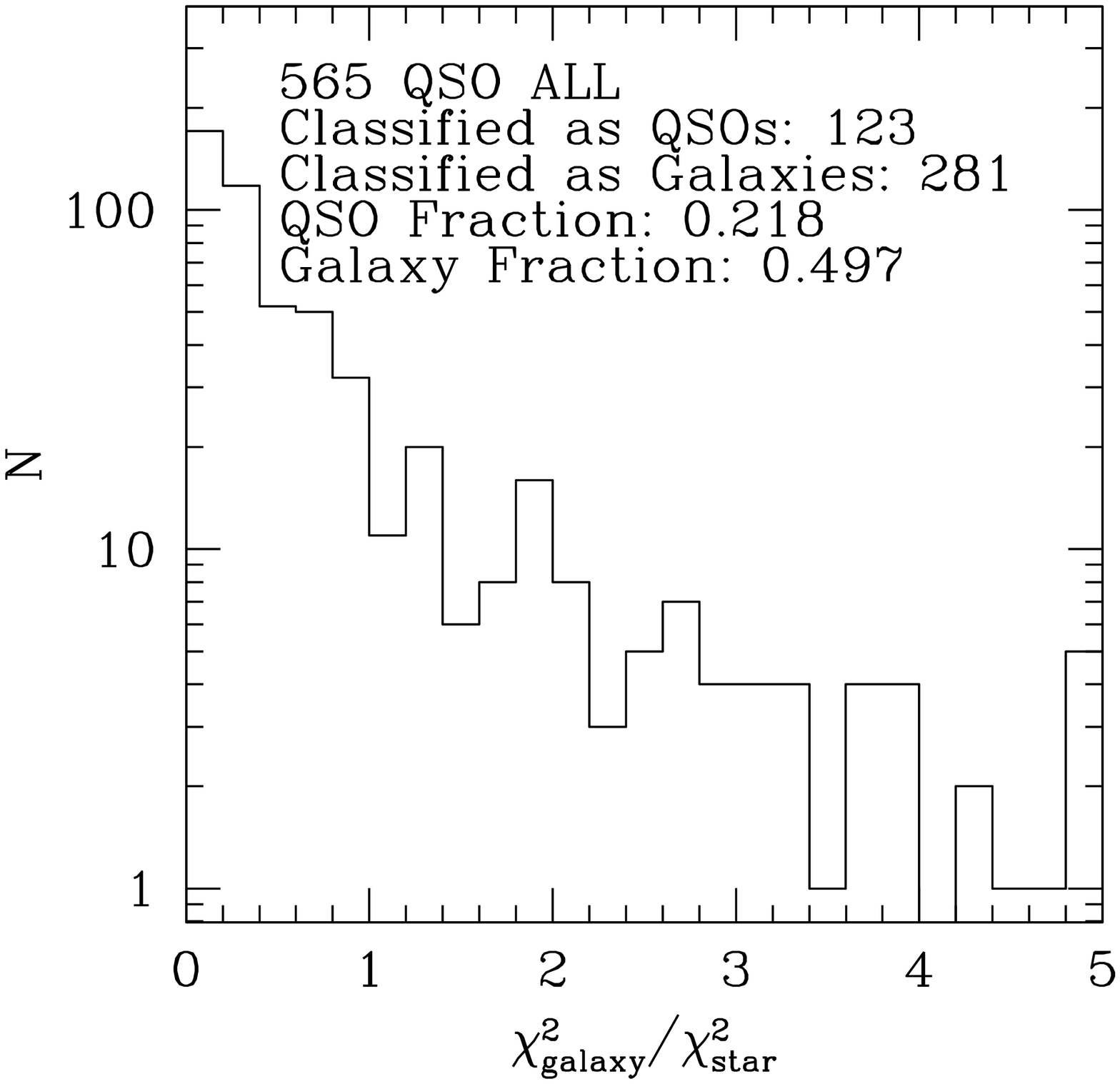}
\includegraphics[width=8cm]{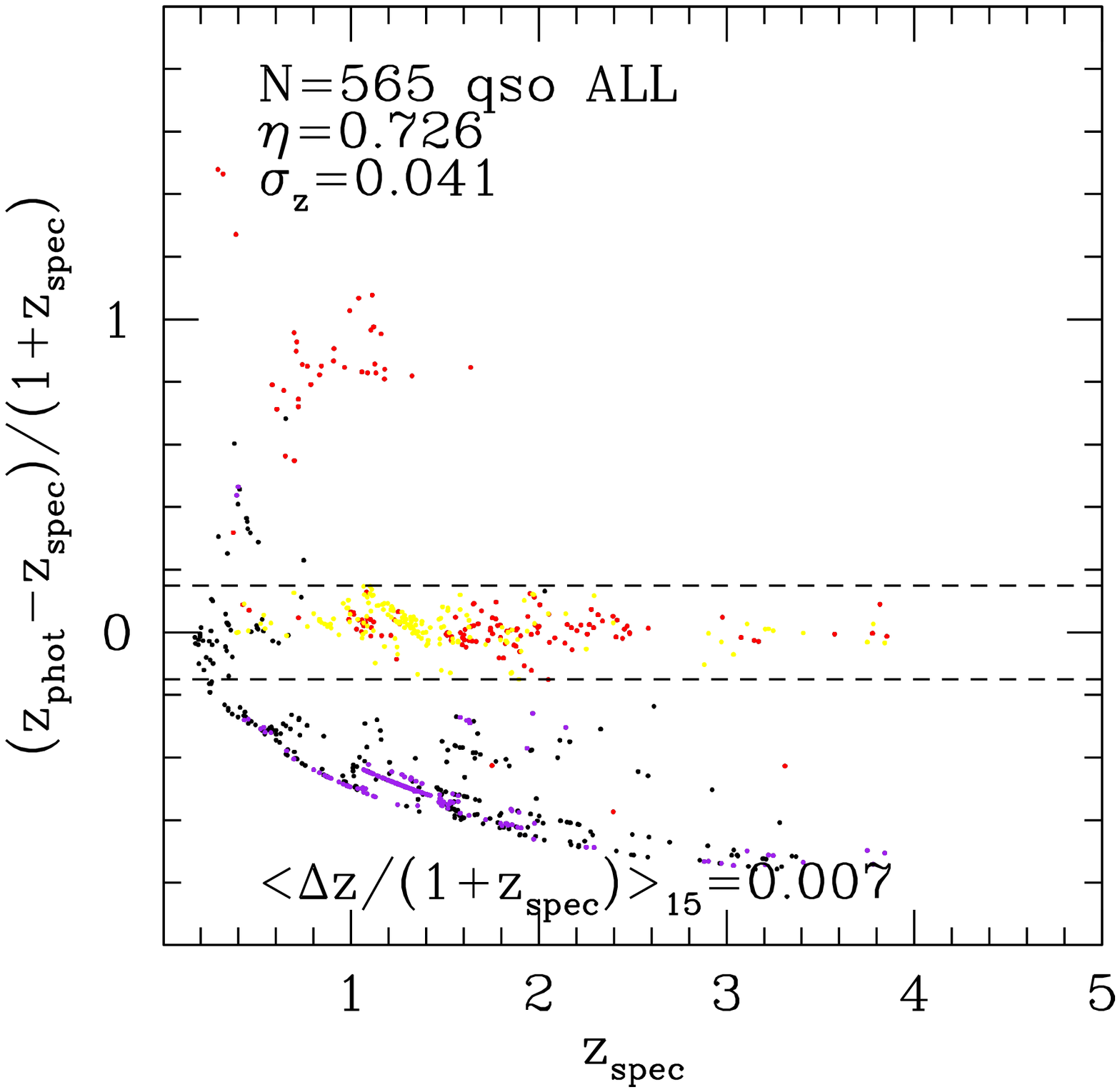}
}
\caption{PanZ performances for QSO. Left: for 29\% of
  spectroscopically confirmed SDSS QSOs PanZ finds a stellar SED as
  the best to the \PS\ photometry
  (i.e. $\chi^2_{galaxy}>\chi^2_{star}$, note that for plotting
  convenience $\chi^2$ ratios are shown). Right: PanZ redshifts for
  QSOs. The red dots show the cases where the QSO SED gives the
  best-fit. The yellow dots show the second-best fit, given by the QSO
  SED, in cases of catastrophic failures (purple points), where a good
  redshift is obtained.}
\label{fig:PanZSLOANQSO}
\end{center}
\end{figure}

Finally, we consider the class of QSOs. Fig. \ref{fig:PanZSLOANQSO},
left, shows that spectroscopically confirmed SDSS QSOs are classified
as QSOs (i.e. are best fit by the QSO SED) in 22\% and as galaxies in
50\% of the cases. As expected, the photometric redshifts are very
poor (Fig.  \ref{fig:PanZSLOANQSO}, right). The QSO SED in the sample
is selected as giving the best-fit in 27\% of the cases, giving the
right redshift in 20\% of the cases. For an additional 28\% where
catastrophically wrong redshifts are derived, the QSO SED gives the
second best fit and a reasonable redshift. Still, if we allow only the
QSO SED to be used, we get a good redshift ($\approx 5$\% in
$\sigma_z$) for only 61\% of the objects. We are in the process of
adding some more QSO SEDs to model better the redshift dependence of
QSO evolution. First tests show that only modest improvements can be
achieved, since we are hitting the intrinsic limitations of the \PS\
filter photometry, combined with the well known difficulties of
deriving photometric redshifts for the power-law like, feature-weak
shape of QSO SEDs \citep{Budavari01, Salvato11}.  
The addition of the u band
certainly improves the results a lot. When we derive photometric
redshifts using the SDSS ugriz Petrosian magnitudes, we get a best-fit
with the QSO SED in 51\% of the cases (with a photometric redshift
good to 5\% in 43\% of the cases), and for an additional 19\% the QSO
SED gives the second best solution with the correct redshift.  If we
allow only the QSO SED to be used, we get a good redshift ($\approx
5$\% in $\sigma_z$) for 70\% of the objects.

Table \ref{table:confusionmatrixPanZ} shows the confusion matrix for
PanZ as a Star/QSO/Galaxy photometric classifier. PanZ performs as
well as PanDiSC when classifying galaxies, but is poorer when it comes
to stars and QSOs, probably due to a lack of appropriate SED
templates. As a consequence, the false positive contamination is
higher for stars (53\%) and galaxies (8\%) classes, but lower (4\%)
for QSOs, compared to PanDiSC. Finally, it is interesting to note
that PanZ biases the classification in a different way than PanDiSC:
there are 29 stars and 3 QSOs recognized as such by PanZ but
not by PanDiSC.

Finally, Table \ref{table:confusionmatrixPanZsdss} shows the confusion
matrix for PanZ as a Star/QSO/Galaxy photometric classifier when the
SDSS Petrosian ugriz photometry is used. The percentage of correctly
classified QSOs doubles (but is still not competitive with the results
of PanDiSC) to reach 44\%, the star classification is slightly
improved to 80\%  and the success in the galaxy
classification is slightly worse (85\%). Therefore, the presence of
the u band helps in the classification of (blue) stars and quasars,
but does not compensate the absence of the y and of good z band data
for galaxies.

As discussed in Sect. \ref{sec:testPanDiSC}, the final assessment of
the relative performances of PanDiSC and PanZ as morphological
classifiers will be made when larger \PS\ catalogs will allow the
derivation of optimal probability thresholds.

\begin{table}[htdp]
  \caption{PanZ as a Star/QSO/Galaxy photometric classifier: 
the confusion matrix in fractions normalized to 1
with true classes in rows. }
\begin{center}
\begin{tabular}{lrrr}
\hline
\hline
True classes & Star & Galaxy & Quasar \\
\hline
Star         & 0.730 & 0.241 & 0.029  \\
Galaxy       & 0.017 & 0.981 & 0.002  \\
QSO          & 0.285 & 0.497 & 0.218  \\
\hline
\end{tabular}
\end{center}
\label{table:confusionmatrixPanZ}
\end{table}

\begin{table}[htdp]
  \caption{PanZ as a Star/QSO/Galaxy photometric classifier using the SDSS Petrosian ugriz photometry: 
    the confusion matrix in fractions normalized to 1
    with true classes in rows. }
\begin{center}
\begin{tabular}{lrrr}
\hline
\hline
True classes & Star & Galaxy & Quasar \\
\hline
Star         & 0.797 & 0.166 & 0.036  \\
Galaxy       & 0.131 & 0.849 & 0.020  \\
QSO          & 0.037 & 0.522 & 0.441  \\
\hline
\end{tabular}
\end{center}
\label{table:confusionmatrixPanZsdss}
\end{table}

\section{Conclusions}
\label{sec:conclusions}

We presented the Photometric Classification Server of \PS , a
database-suppor\-ted, fully automatised package to classify \PS\
objects into stars, galaxies and quasars based on their \PS\ colors
and compute the photometric redshifts of extragalactic objects.  Using
the high signal-to-noise photometric catalogs derived for the \PS\
Medium-Deep Fields we provide preliminary Star/QSO/Galaxy
classifications and demonstrate that excellent photometric redshifts
can be derived for the sample of Luminous Red Galaxies. Further tuning
of our probabilistic classifier with the large \PS\ catalogs available
in the future will optimize its already nice performances in terms of
completeness and purity.  Applied to the photometry that the $3\pi$
survey is going to deliver, possibly combined with u-band or
near-infrared photometry coming from other surveys, this will allow us
to build up an unprecedented large sample of LRGs with accurate
distances. In a future development of PCS we will include size and/or
morphological information to improve further the object
classification, implement the PanSTeP (\PS\ Stellar Parametrizer)
software to constrain stellar parameters, and enlarge the SED sample
to follow LRGs to higher redshifts and possibly improve results for
blue galaxies and QSOs. Alternative photometric redshift codes could
also be considered.  Moreover, the independent classification
information coming from PanZ and PanDiSC will be merged and used to
iterate on the photometric redshifts, by narrowing down the choice of
SEDs, or deciding which photometry (psf photometry for point-objects
versus extended sources photometry for galaxies) is more appropriate
for each object.

{\it Facilities:} \facility{PS1 (GPC1)}

\acknowledgments


The Pan-STARRS1 Survey has been made possible through contributions of
the Institute for Astronomy, the University of Hawaii, the Pan-STARRS
Project Office, the Max-Planck Society and its participating
institutes, the Max Planck Institute for Astronomy, Heidelberg and the
Max Planck Institute for Extraterrestrial Physics, Garching, The Johns
Hopkins University, Durham University, the University of Edinburgh,
Queen's University Belfast, the Harvard-Smithsonian Center for
Astrophysics, and the Las Cumbres Observatory Global Telescope
Network, Incorporated, the National Central University of Taiwan, and
the National Aeronautics and Space Administration under Grant
No. NNX08AR22G issued through the Planetary Science Division of the
NASA Science Mission Directorate. Partial support for this work was
provided by National Science Foundation grant AST-1009749.

\clearpage

\end{document}